\documentstyle[preprint,aps,epsf]{revtex}

\newcommand{\be}{\begin{equation}}
\newcommand{\ba}{\begin{eqnarray}}
\newcommand{\ee}{\end{equation}}
\newcommand{\ea}{\end{eqnarray}}
\newcommand{\nn}{\nonumber}
\def\prd#1#2#3{{\it Phys.~Rev.\/}~{\bf D#1} (19#2) #3}
\def\np#1#2#3{{\it Nucl.~Phys.\/}~{\bf B#1} (19#2) #3}
\def\pl#1#2#3{{\it Phys.~Lett.\/}~{\bf B#1} (19#2) #3}

\begin{document}

\preprint{\setlength{\baselineskip}{1.5em}\vbox{\hbox{PITHA 98/2}}}

\draft

\title{Magnetic and axial vector form factors as probes of \\
orbital angular momentum in the proton}

\author{M. Casu\footnote{email: casu@physik.rwth-aachen.de} and
L.~M.~Sehgal\footnote{email: sehgal@physik.rwth-aachen.de}}
\address{Institut f\"ur Theoretische Physik (E), RWTH Aachen\\
D-52056 Aachen, Germany}

\maketitle


\begin{abstract}
We have recently examined the static properties of the baryon octet
(magnetic moments and axial vector coupling constants) in a generalized
quark model in which the angular momentum of a polarized nucleon is 
partly spin $\langle S_z \rangle$ and partly orbital $\langle L_z \rangle$.
The orbital momentum was represented by the rotation of a flux-tube 
connecting the three constituent quarks. The best fit is obtained with 
$\langle S_z \rangle = 0.08\pm 0.15$, $\langle L_z \rangle = 0.42\pm 0.14$.
We now consider the consequences of this idea for the $q^2$-dependence 
of the magnetic and axial vector form factors. It is found that the 
isovector magnetic form factor $G_M^{\mathrm{isovec}}(q^2)$ differs in shape 
from the axial form factor $F_A(q^2)$ by an amount that depends on the
spatial distribution of orbital angular momentum. The model of a rigidly 
rotating flux-tube leads to a relation between the magnetic, axial vector 
and matter radii, $\langle r^2 \rangle_{\mathrm{mag}} = f_{\mathrm{spin}} \langle 
r^2 \rangle_{\mathrm{axial}} + \frac{5}{2} f_{\mathrm{orb}} \langle r^2 \rangle
_{\mathrm{matt}}$, where $f_{\mathrm{orb}}/f_{\mathrm{spin}} = \frac{1}{3}\langle 
L_z \rangle / G_A$, $f_{\mathrm{spin}} + f_{\mathrm{orb}} = 1$. The shape of 
$F_A(q^2)$ is found to be close to a dipole with $M_A = 0.92\pm 0.06$ GeV. 
\end{abstract}



\section{Introduction}
In a recent paper \cite{Casu} we performed a fit to the magnetic moments of 
the baryon octet in a model in which these quantities are determined partly
by the quark spins $\Delta u$, $\Delta d$, $\Delta s$, and partly by an
orbital angular momentum $\langle L_z \rangle$, shared between the constituent
quarks. The model is exemplified by the following ansatz for the proton and
neutron magnetic moments:
\parbox{15cm}{\begin{eqnarray*}
\mu_p &=& \mu_u \Delta u + \mu_d \Delta d + \mu_s \Delta s  + \left [ 
\frac{2}{3}\mu_u + \frac{1}{3}\mu_d \right ] \langle L_z \rangle, \\
\mu_n &=& \mu_u \Delta d + \mu_d \Delta u + \mu_s \Delta s + \left [ 
\frac{1}{3}\mu_u + \frac{2}{3}\mu_d \right ] \langle L_z \rangle.
\end{eqnarray*}}\hfill
\parbox{1cm}{\ba\label{mupmun}\ea}
The part containing $\Delta u$, $\Delta d$, $\Delta s$, is the ``spin'' 
contribution to the magnetic moments, arising from the polarization 
of quarks and antiquarks in a polarized proton:
\be
\Delta q = ( q_+ - q_- ) + ( \bar{q}_+ - \bar{q}_- ).
\ee
The part proportional to $\langle L_z \rangle$ is the ``orbital'' (or
convective) contribution, determined by the prescription of dividing 
the orbital angular momentum in proportion to the constituent quark 
masses. The complete set of baryon magnetic moments obtained by this 
prescription is shown in table \ref{tab1}. These expressions, without the 
orbital part, were written down in Refs. \cite{Karl,Bartelski}.

\begin{figure}[b]
\begin{center}
\mbox{\epsfysize 5cm \epsffile{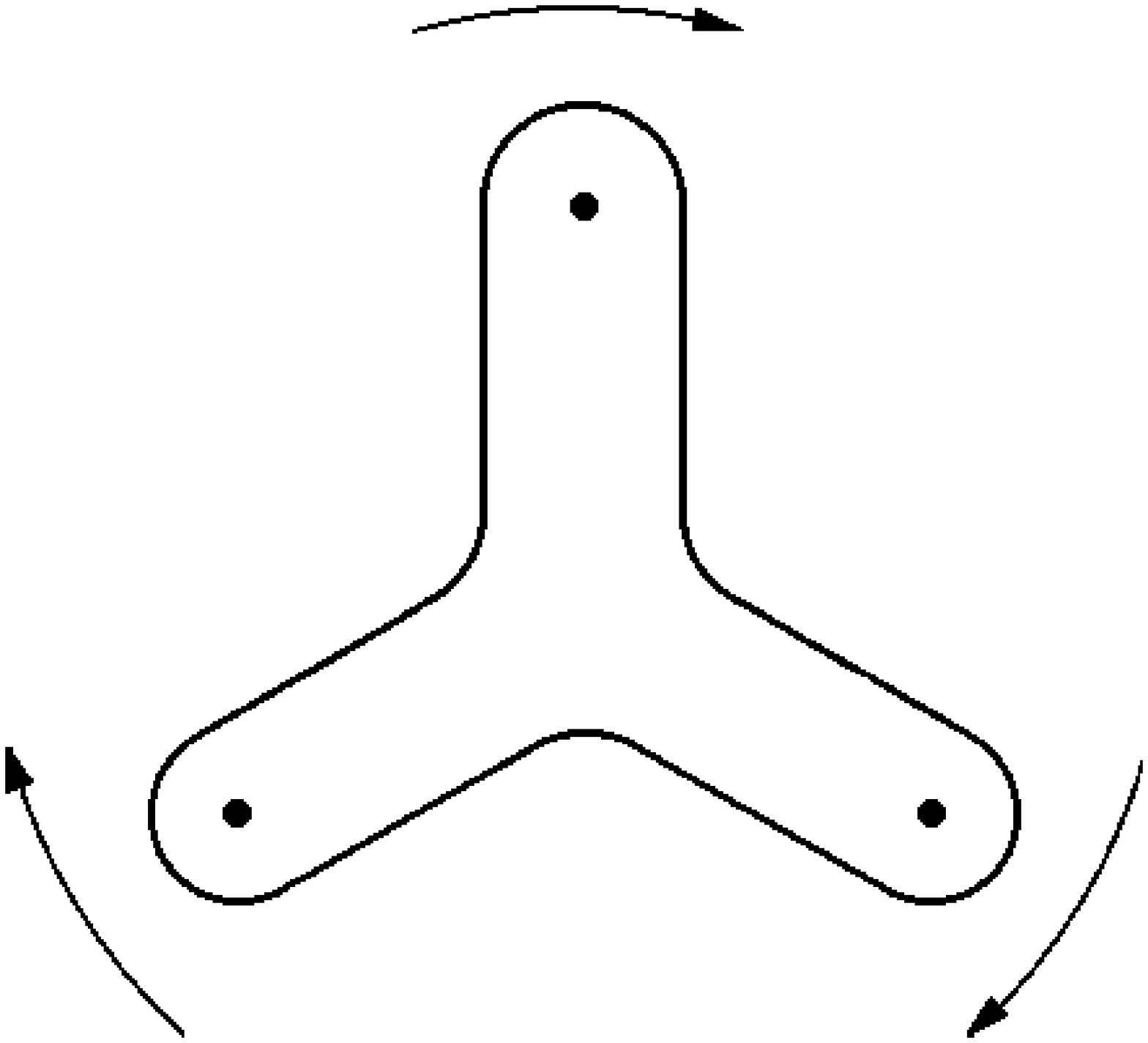}}
\end{center}
\caption{}
\label{rot}
\end{figure}

There are two essential elements that go into the above equations for 
the magnetic moments:
\begin{itemize}
\item[(1)] It is assumed that one may use the quark spins
$\Delta q$ in place of the quantities
\be
\delta q = ( q_+ - q_- ) - ( \bar{q}_+ - \bar{q}_- )
\ee
that are appropriate to an expression for the magnetic moment. This 
approximation is justified if antiquarks in a proton carry little 
polarization. An example of such a situation is the chiral quark model 
\cite{Cheng}, in which antiquarks are embedded in a cloud of spin-zero
mesons. 
\item[(2)] The partition of $\langle L_z \rangle$ in proportion to the 
masses of the constituent quarks is based on the picture of a baryon as
a symmetric three-pronged flux-tube of equal segments (fig.\ref{rot}), rotating 
collectively around the spin axis \cite{Casu}. The appearence of the 
same magnetons $\mu_u$, $\mu_d$, $\mu_s$, in the orbital as in the spin part
means, in particular, that the orbital $g$-factor has been taken to be
$g_l=1$. (In a more general description, one could interpret $\langle 
L_z \rangle$ as $\langle g_l L_z \rangle$.)
\end{itemize}


\section{$S_z$ and $L_z$ from static properties}
\begin{figure}[h]  
\begin{center}
\mbox{\epsfysize 17cm \epsffile{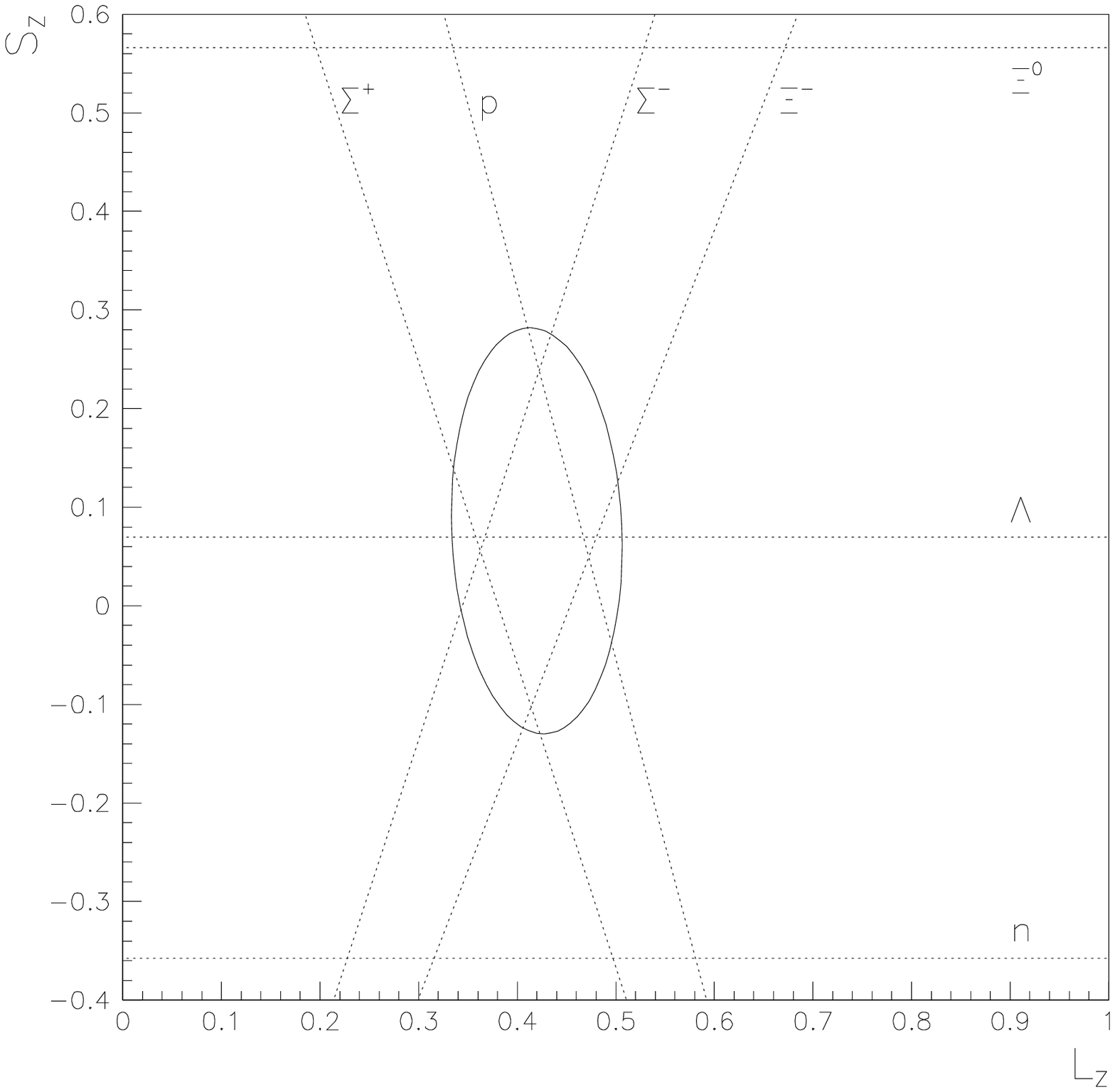}}
\end{center}
\caption{}
\label{szlz}
\end{figure}

\begin{figure}[b]
\begin{center}
\mbox{\epsfysize 17cm \epsffile{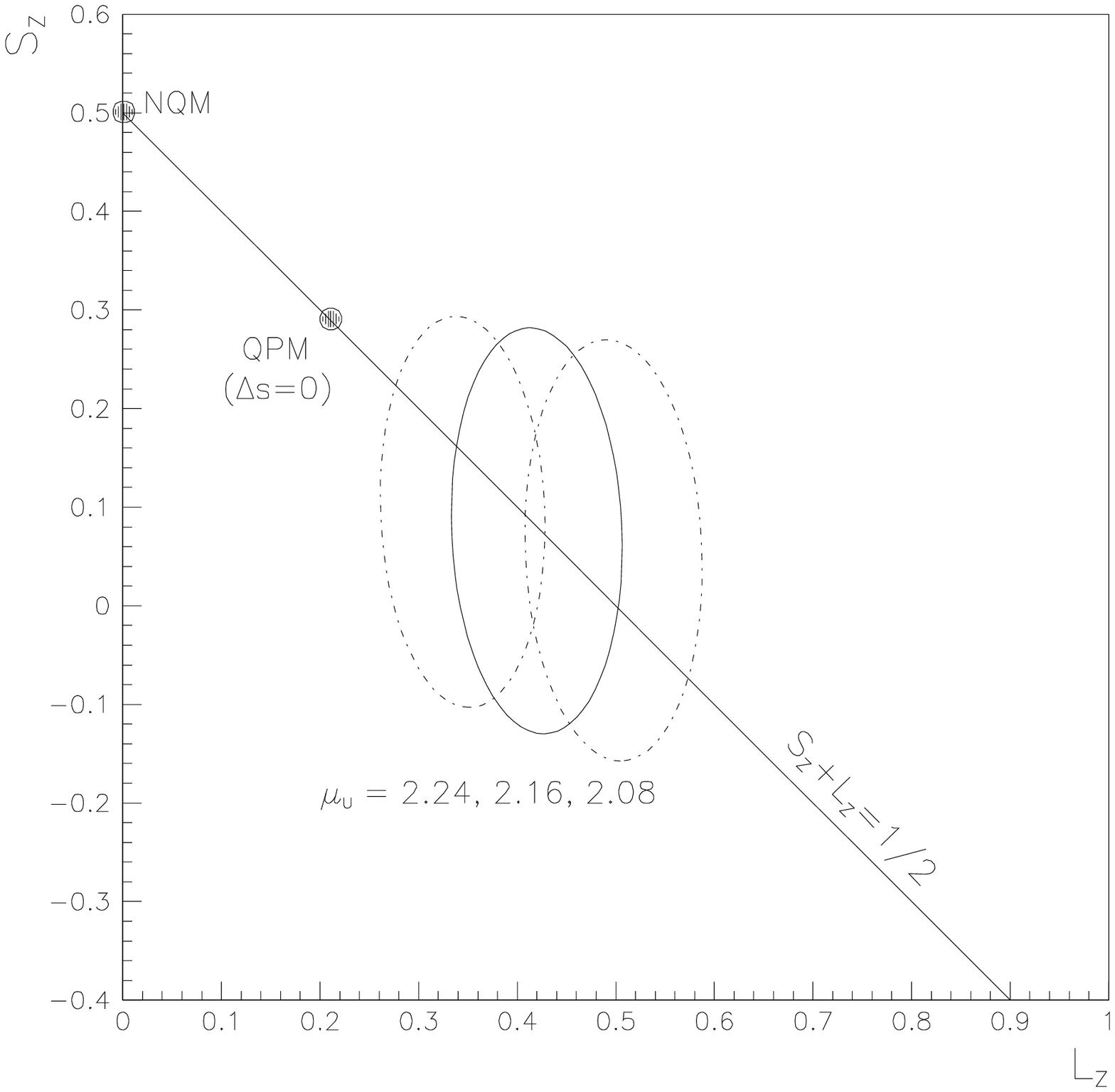}}
\end{center}
\caption{}  
\label{szlzerrors}
\end{figure}

The fit to the empirical values of the magnetic moments in \cite{Casu} 
was carried out under the following constraints:
\begin{itemize}
\item[(i)] The quark magnetic moments were assumed to satisfy $\mu_u = 
-2\mu_d$, $\mu_s = 0.6 \mu_d$.
\item[(ii)] The quark spins $\Delta u$, $\Delta d$ and $\Delta s$  were 
constrained to satisfy the measured values of the axial vector couplings
$a^{(3)}$ and $a^{(8)}$:\\
\parbox{13cm}{\begin{eqnarray*}
a^{(3)} &=& \Delta u - \Delta d = 1.26 \\ 
a^{(8)} &=& \Delta u + \Delta d - 2\Delta s = 0.58
\end{eqnarray*}}\hfill
\parbox{1cm}{\ba\ea}
These conditions are equivalent to the statement $F=0.46$, $D=0.80$, in terms
of which $a^{(3)}=F+D$ ($\equiv G_A$, the axial vector coupling constant 
of neutron decay) and $a^{(8)}=3F-D$.
\item[(iii)] Each magnetic moment was assigned a theoretical uncertainty of
$\pm 0.1 \mu_N$ (as in Ref.\cite{Karl}). This ensured that all of the baryons 
were given essentially the same weight in the fit and the $\chi^2$ per degree 
of freedom was about unity.
\end{itemize}
In this manner, the magnetic moments are reduced to functions of three 
variables, which we choose to be $\mu_u$, $\langle L_z \rangle$ and $\langle
S_z \rangle$, the last being defined as 
\be
\langle S_z \rangle = \frac{1}{2} \left ( \Delta u + \Delta d + \Delta s 
\right ) \equiv \frac{1}{2} \Delta\Sigma.
\ee 
The result of the fit is
\be
\label{fitresult}
\mu_u = 2.16 \pm 0.08,\qquad \langle S_z \rangle = 0.076 \pm 0.13,\qquad
\langle L_z \rangle = 0.42 \pm 0.10
\ee
with $\chi^2$/DOF = 1.1. For the central value of $\mu_u$, the allowed domain 
of $\langle S_z \rangle$ and $\langle L_z \rangle$ is given by the ellipse
shown in fig.\ref{szlz}. Allowing $\mu_u$ to vary over the interval given
in eq.(\ref{fitresult}), we obtain the domain shown in fig.\ref{szlzerrors}, 
from which we infer a final estimate 
\be
\label{szlzfit}
\langle S_z \rangle = 0.08 \pm 0.15,\qquad \langle L_z \rangle = 0.42 \pm 0.14.
\ee
It is remarkable that the domain of $\langle S_z \rangle$ and $\langle L_z 
\rangle$ determined by the static properties of the baryons fulfils rather
closely the condition $\langle S_z \rangle + \langle L_z \rangle = \frac{1}{2}$.
That is, the spin and orbital momenta of the quarks and antiquarks saturate 
the angular momentum of the proton, {\it without imposing this as an external
requirement}. This may be regarded as {\it a posteriori} justification for
the assumption $g_l=1$. The fact that $\langle S_z \rangle + \langle L_z \rangle
\approx \frac{1}{2}$ supports the idea, that the spin
and orbital angular momentum are linked together by a transition of the form
$q_+ \to q'_-+M$ ($L=1$), $M$ being a spin-zero meson \cite{Cheng}. This in turn
provides support to the assumption $\delta q \approx \Delta q$, 
based on negligible antiquark polarization.

Also indicated in fig.\ref{szlzerrors} is the location of two ``sign-posts'', 
that serve as reference points in the angular momentum structure:
\begin{itemize}
\item[(i)] NQM: This is the ``naive quark model'', which describes the 
nucleon as 3 independent quarks in $1s$ orbits, corresponding to $\langle S_z 
\rangle = \frac{1}{2}$, $\langle L_z \rangle = 0$. The $SU(6)$ symmetry of the 
model leads to the prediction $\Delta u = \frac{4}{3}$, $\Delta d =-\frac{1}{3}$,
$\Delta s =0$, axial vector couplings $a^{(3)}= \frac{5}{3}$, $a^{(8)} = 1$,
and the magnetic moment ratio $\mu_p/\mu_n = -\frac{3}{2}$.
\item[(ii)] QPM ($\Delta s = 0$): This is the special case of the quark
parton model discussed in Ref. \cite{Sehgal1}, in which $\Delta u$ and $\Delta d$
were allowed to be free, but $\Delta s$ was neglected. The characteristic 
prediction of this model is $a^{(8)} = a^{(0)}$, where $a^{(0)}=\Delta u + 
\Delta d + \Delta s$,  implying $\langle S_z \rangle =
\frac{1}{2}a^{(0)} = \frac{1}{2}(3F-D) = 0.29$, the
remaining angular momentum being attributed to $\langle L_z \rangle = 
\frac{1}{2} -  \langle S_z \rangle = 0.21$. This version of the QPM leads to
the Ellis-Jaffe sum rules \cite{Ellis} for polarized structure functions:
\be
\int g_1^p(x)dx = \frac{1}{2}\left ( F - \frac{1}{9} D \right ),\qquad
\int g_1^n(x)dx = \frac{1}{3}\left ( F - \frac{2}{3} D \right ).
\ee
\hspace*{-1.1cm}
\end{itemize}

In what follows, we consider a test for the presence of orbital angular
momentum $\langle L_z \rangle$ and its specific association with the 
collective rotation of the constituent quarks. 


\section{Tests for $L_z$ in magnetic and axial vector form factors}
\begin{figure}[b]
\begin{center}
\mbox{\epsfysize 17cm \epsffile{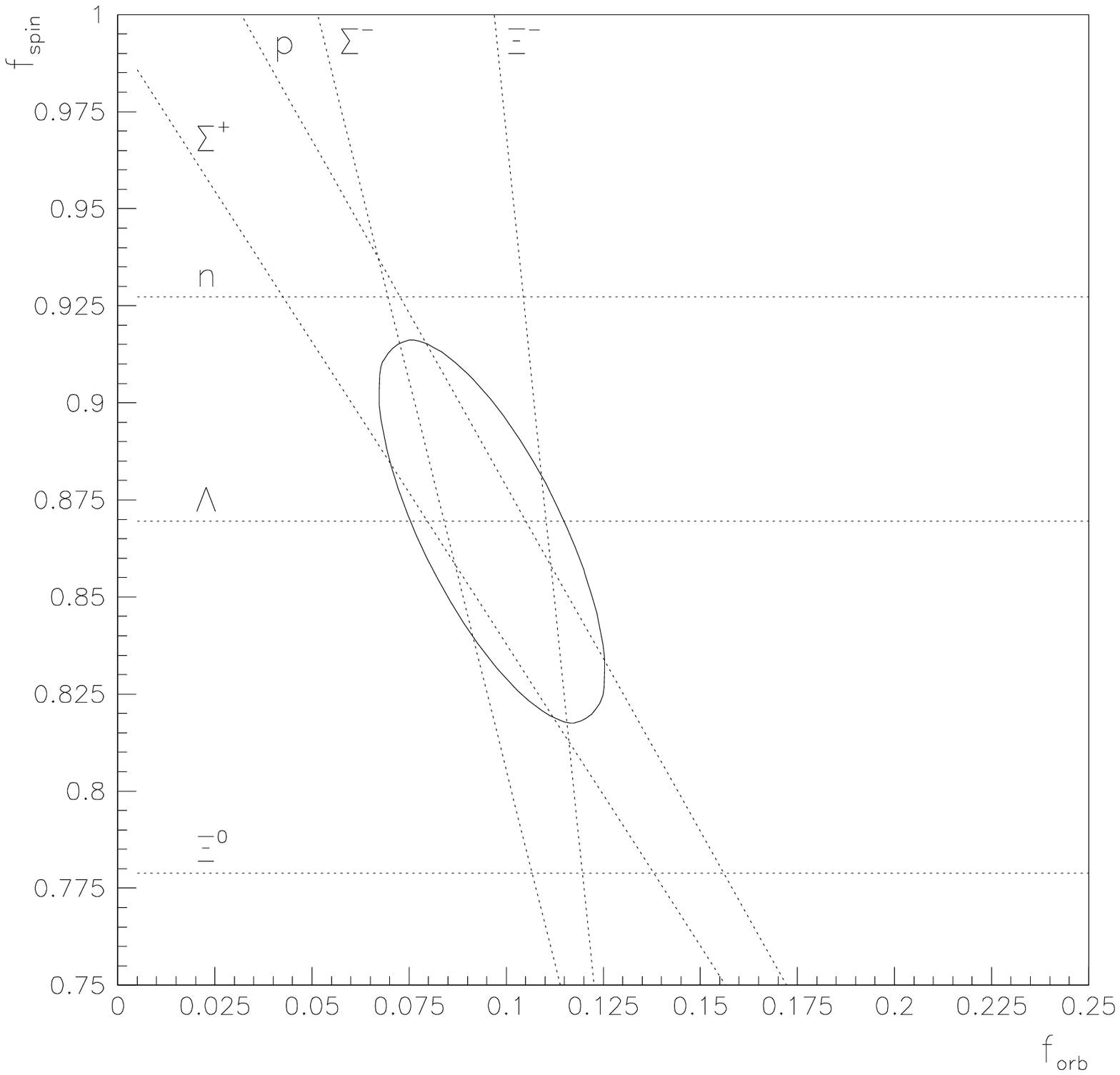}}
\end{center}
\caption{}
\label{fspinforb}
\end{figure}

We focus on the isovector magnetic moment of the nucleon, obtained
by taking the difference of $\mu_p$ and $\mu_n$ in eq.(\ref{mupmun}):
\be
\label{isovec}
\mu_p - \mu_n = \left ( \mu_u - \mu_d \right ) \left [ G_A + \frac{1}{3}
\langle L_z \rangle \right ].
\ee
Note that the terms containing $\Delta s$ cancel in the difference.
We regard this equation as a decomposition of the isovector magnetic moment 
into a part depending on the axial vector charge and a part depending on 
orbital angular momentum. Introducing the abbreviation
\be
f_{\mathrm{spin}} \equiv \frac{\mu_u - \mu_d}{\mu_p - \mu_n} G_A,\qquad
f_{\mathrm{orb}} \equiv \frac{\mu_u - \mu_d}{\mu_p - \mu_n} \frac{1}{3}
\langle L_z \rangle,
\ee
eq.(\ref{isovec}) amounts to 
\be
\label{fspin_forb}
1 = f_{\mathrm{spin}} + f_{\mathrm{orb}}.
\ee
Returning to the three-parameter fit given by eq.(\ref{fitresult}), we 
can regard the fitted parameters as being $\langle S_z \rangle$, 
$f_{\mathrm{spin}}$ and $f_{\mathrm{orb}}$ (in place of $\langle S_z \rangle$, 
$\mu_u$ and $\langle L_z \rangle$). For the central value of 
$\langle S_z \rangle$, the domain of $f_{\mathrm{spin}}$ and $f_{\mathrm{orb}}$
determined by the various magnetic moments is shown in fig. \ref{fspinforb}. 
The fitted values, taking into account the spread
of $\langle S_z \rangle$, are $f_{\mathrm{spin}} = 0.87 \pm 0.03$, 
$f_{\mathrm{orb}} = 0.096 \pm 0.03$.
Considering that these values nearly satisfy the isovector magnetic 
moment relation, $f_{\mathrm{spin}} + f_{\mathrm{orb}} = 1$, we use the
following approximate values, which fulfil eq.(\ref{fspin_forb}) 
exactly
\be
\label{fspin_forb_approx}
f_{\mathrm{spin}} = 0.90 \pm 0.03,\qquad f_{\mathrm{orb}} = 0.10 \pm 0.03.
\ee
Note that the ratio $f_{\mathrm{orb}}/f_{\mathrm{spin}} = \frac{\langle L_z 
\rangle}{3G_A} = \frac{1}{9}$ implies $\langle L_z \rangle = 0.42$, 
as given in eq.(\ref{fitresult}). Eq.(\ref{fspin_forb_approx}) amounts
to the statement, that the isovector magnetic moment is 90\% due to quark
spin polarization and 10\% due to quark rotation. 

We now define spatial 
distributions $\rho_{\mathrm{mag}}(r)$, $\rho_{\mathrm{axial}}(r)$ and 
$\rho_{\mathrm{orb}}(r)$ whose volume integrals yield the quantities ($\mu_p 
- \mu_n$), $G_A$ and $\langle L_z \rangle$ appearing in eq.(\ref{isovec}):
\ba
\mu_p - \mu_n &=& \int d^3x\,\rho_{\mathrm{mag}}(r), \nn\\
G_A &=& \int d^3x\,\rho_{\mathrm{axial}}(r), \\
\langle L_z \rangle &=& \int d^3x\,\rho_{\mathrm{orb}}(r) \nn .
\ea
The local form of eq.(\ref{isovec}) then reads
\be
\label{rho}
\rho_{\mathrm{mag}}(r) = \left ( \mu_u - \mu_d \right ) \left [ 
\rho_{\mathrm{axial}}(r)+ \frac{1}{3}\rho_{\mathrm{orb}}(r)\right ].
\ee
Introducing, for convenience, `` normalized'' densities
\ba
\tilde{\rho}_{\mathrm{mag}}(r) &\equiv& \rho_{\mathrm{mag}}(r) / ( \mu_p - 
\mu_n ), \nn\\
\tilde{\rho}_{\mathrm{axial}}(r) &\equiv& \rho_{\mathrm{axial}}(r) / G_A, \\
\tilde{\rho}_{\mathrm{orb}}(r) &\equiv& \rho_{\mathrm{orb}}(r) / \langle L_z 
\rangle, \nn
\ea
eq.(\ref{rho}) assumes the form
\be
\label{rhotilde}
\tilde{\rho}_{\mathrm{mag}}(r) = f_{\mathrm{spin}}\,\tilde{\rho}
_{\mathrm{axial}}(r) + f_{\mathrm{orb}}\,\tilde{\rho}_{\mathrm{orb}}(r).
\ee
The functions $\tilde{\rho}_{\mathrm{i}}(r)$ all satisfy $\int\tilde{\rho}
_{\mathrm{i}}(r)d^3x = 1$, so that the integrated form of eq.(\ref{rhotilde}) 
is simply the relation (\ref{fspin_forb}). Fourier transforming 
eq.(\ref{rhotilde}), we get a relation between the isovector magnetic, axial 
vector and ``orbital'' form factors of the nucleon:
\be
\label{isoff}
H_{\mathrm{mag}}^{\mathrm{isovec}}(Q^2) = f_{\mathrm{spin}}\,H_{\mathrm{axial}}
(Q^2) + f_{\mathrm{orb}}\,H_{\mathrm{orb}}(Q^2)
\ee
where 
\be
H_{\mathrm{i}}(Q^2=\vec{Q}^2) =\int\,\tilde{\rho}_{\mathrm{i}}(r)\,e^{i\vec{Q}
\vec{x}}d^3x
\ee
with $H_{\mathrm{i}}(0) =1$.

The form factor $H_{\mathrm{mag}}^{\mathrm{isovec}}(Q^2)$ is an experimentally 
measured quantity, related to the magnetic (Sachs) form factors of the proton 
and the neutron by
\be
H_{\mathrm{mag}}^{\mathrm{isovec}}(Q^2) = \frac{G_M^p(Q^2) - G_M^n(Q^2)}{\mu_p 
- \mu_n}.
\ee
To the extent that $ G_M^p(Q^2)$ and $G_M^n(Q^2)$ are both proportional 
to $(1+Q^2/0.71\,\mathrm{GeV}^2)^{-2}$, we have 
\be
\label{magiso}
H_{\mathrm{mag}}^{\mathrm{isovec}}(Q^2) = \frac{1}{\left ( 1 + \frac{Q^2}{M_V^2} 
\right )^2},\qquad M_V = 0.84 \mathrm{GeV}.
\ee
The (normalized) axial vector form factor is likewise usually parametrized 
as a dipole
\be
\label{dipole}
H_{\mathrm{axial}}(Q^2) = \frac{1}{\left ( 1 + \frac{Q^2}{M_A^2} \right )^2}.
\ee
It is clear from eq.(\ref{isoff}), that the difference between 
$H_{\mathrm{mag}}^{\mathrm{isovec}}(Q^2)$ and $H_{\mathrm{axial}}(Q^2)$ is a 
measure of the orbital contribution proportional to $f_{\mathrm{orb}}$: in the 
limit $f_{\mathrm{orb}}=0$, $f_{\mathrm{spin}}=1$, these two form factors would 
be identical and we would have $M_A\equiv M_V$.

The orbital form factor $H_{\mathrm{orb}}(Q^2)$ is a calculable feature of our
model, which ascribes the orbital angular momentum to the rigid rotation
of a flux-tube. Assuming matter in the proton to be distributed as 
$\rho_{\mathrm{matt}} \propto e^{-r/a}$, the density of orbital angular momentum 
$\tilde{\rho}_{\mathrm{orb}}$ is proportional to $r^2e^{-r/a}$.
The resulting orbital form factor is
\be
\label{horb}
H_{\mathrm{orb}}(Q^2=\vec{Q}^2) = \frac{\int e^{i\vec{Q}\vec{x}}\,r^2\,e^{-r/a}\,d^3x}
{\int r^2\,e^{-r/a}\,d^3x} = \frac{1 - Q^2 a^2}{\left ( 1 + Q^2 a^2 \right )^4}. 
\ee
In particular, the rms radius associated with the orbital form factor is
\be
\langle r^2 \rangle_{\mathrm{orb}} = \frac{30}{a^2}.
\ee
This is to be compared with the rms radius of the matter distribution 
\be
\langle r^2 \rangle_{\mathrm{matt}} = \frac{12}{a^2},\qquad \mathrm{i.e.} \quad
\langle r^2 \rangle_{\mathrm{orb}} = \frac{5}{2}\langle r^2 
\rangle_{\mathrm{matt}}.
\ee  
Eq.(\ref{isoff}) thus implies a relation between the mean square radii
of the various form factors:
\ba
\label{rquad}
\langle r^2 \rangle_{\mathrm{mag}}^{\mathrm{isovec}} &=& f_{\mathrm{spin}}\,
\langle r^2 \rangle_{\mathrm{axial}} + f_{\mathrm{orb}}\,\langle r^2 
\rangle_{\mathrm{orb}}\nn\\
&=& f_{\mathrm{spin}}\,\langle r^2 \rangle_{\mathrm{axial}} + 
\frac{5}{2}f_{\mathrm{orb}}\, \langle r^2 \rangle_{\mathrm{matt}}.
\ea
To the extent that the matter radius of the proton is assumed to be the same
as the magnetic radius, we have the prediction 
\be
\label{rquad2}
\langle r^2 \rangle_{\mathrm{axial}}  = \frac{1-\frac{5}{2} f_{\mathrm{ orb}}}
{f_{\mathrm{spin}}}\langle r^2 \rangle_{\mathrm{mag}}.
\ee
Using the values $f_{\mathrm{orb}} = 0.10 \pm 0.03$, $f_{\mathrm{spin}} = 0.90 
\pm 0.03$ obtained from the fits to the magnetic moments, and the dipole 
parametrization given in eqs.(\ref{magiso}) and (\ref{dipole}), the above 
relation yields 
\be
M_A = (1.10 \pm 0.07)\,M_V = 0.92 \pm 0.06\,\mathrm{GeV}
\ee
in quite reasonable agreement with the value $M_A \approx 1.0$ GeV deduced 
from elastic neutrino-nucleon scattering \cite{Sehgal2}. It may be remarked here 
that measurements of elastic $pp$ and $\bar{p}p$ scattering, when interpreted in a 
geometrical model \cite{Ar} tend to give a matter radius slightly larger than 
the charge radius, namely $\sqrt{\langle r^2 \rangle_{\mathrm{matt}}} \approx 
0.89$ fm, as compared to $\sqrt{\langle r^2 \rangle_{\mathrm{charge}}} 
\approx 0.84$ fm. If this difference is taken into account, the prediction for 
$M_A$ obtained from eq.(\ref{rquad2}) increases by about one per cent.

\begin{figure}[b]
\begin{center}
\mbox{\epsfysize 17cm \epsffile{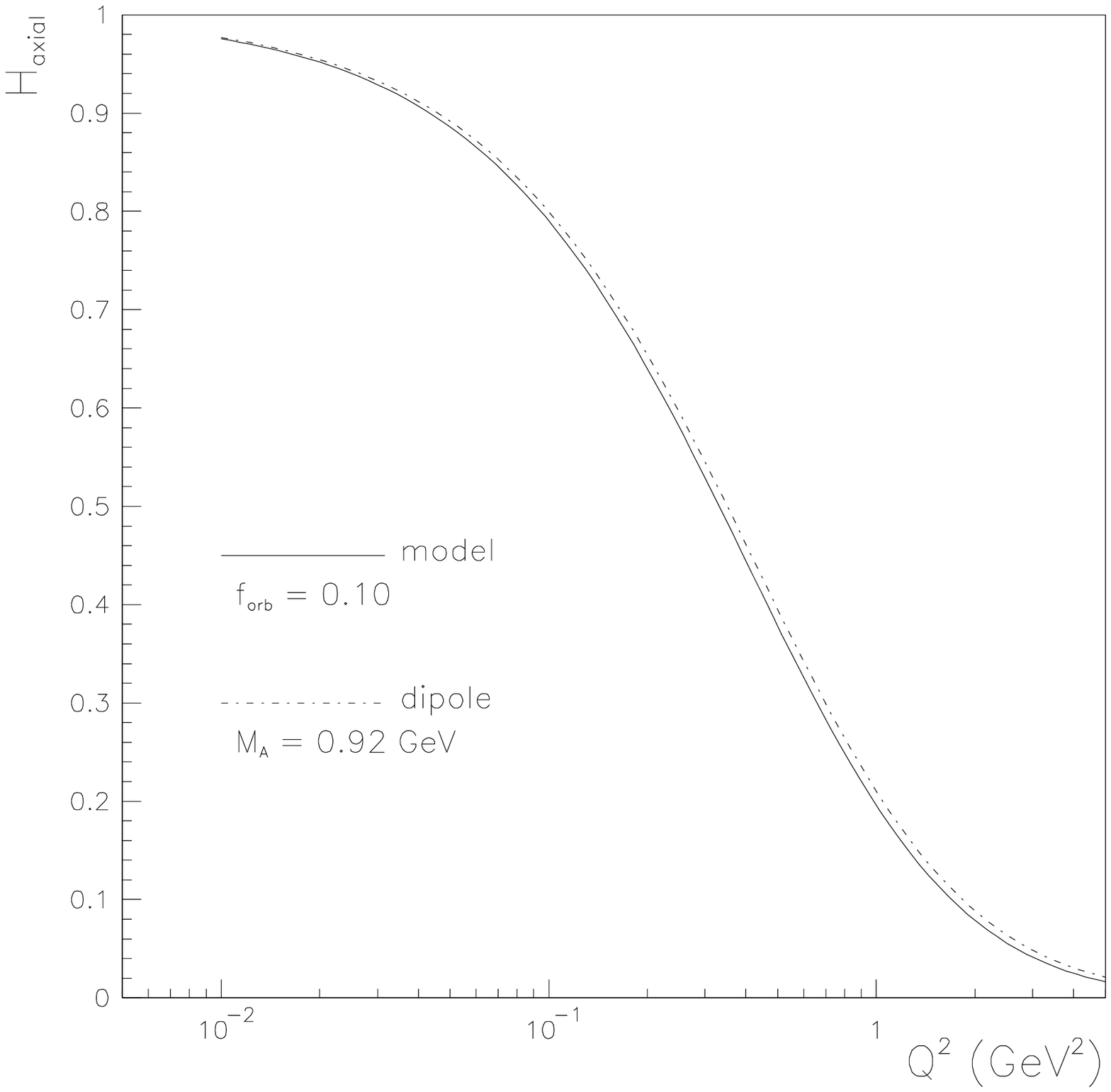}}
\end{center}
\caption{}
\label{haxialplot}
\end{figure}

Finally, we can also obtain from eq.(\ref{isoff}) a more detailed prediction 
for the shape of the axial vector factor $H_{\mathrm{axial}}(Q^2)$, in terms of
the empirically known magnetic form factor $H_{\mathrm{mag}}^{\mathrm{isovec}} = 
(1+Q^2/0.71\,\mathrm{GeV}^2 )^{-2}$ and the calculated orbital form factor 
$H_{\mathrm{orb}}(Q^2)$ given in eq.(\ref{horb}). The result is plotted in 
fig.\ref{haxialplot}, and is close to a dipole with $M_A \approx 0.92$ GeV.


\section{Concluding remarks}
We presented in Ref. \cite{Casu} a model of the proton as a 
collectively rotating system of quarks, with an orbital angular momentum
determined by the baryon magnetic moments and the axial vector couplings
to be $\langle L_z \rangle = 0.42 \pm 0.14$. We have now shown that the 
same assumption of a rigidly rotating structure  leads to a difference 
between the normalized axial vector and isovector magnetic form factors, which is 
dependent on the spatial distribution of orbital angular momentum. The
model of rigid rotation leads to an axial vector form factor which is
close to a dipole with $M_A = 0.92 \pm 0.06$ GeV. Our model of a rotating
matter distribution has some similarity to that discussed by 
Chou and Yang \cite{Chou}, who proposed a test for the velocity profile of a 
polarized proton in hadronic interactions.

It is of interest to ask how our results for $\langle S_z \rangle$ and
$\langle L_z \rangle$, namely
\be
\label{sz_lz_result}
\langle S_z \rangle = 0.08 \pm 0.15,\qquad
\langle L_z \rangle = 0.42 \pm 0.14, 
\ee
compare with those obtained from other considerations. Our fit indicates a
dominance of orbital over spin angular momentum. This feature is opposite  to 
that in the non-relativistic quark model,
\be
\langle S_z \rangle = \frac{1}{2},\quad \langle L_z \rangle = 0, \quad\quad 
(\mathrm{NQM}) 
\ee
and closer to the soliton picture of the proton represented by the Skyrme model
\cite{Brodsky}
\be
\langle S_z \rangle = 0,\quad \langle L_z \rangle = \frac{1}{2}, \quad\quad 
(\mathrm{Skyrme}). 
\ee
An interesting version of the soliton model,that interpolates between the NQM and
Skyrme limits, is the chiral quark soliton picture \cite{Kim}, which predicts
\be
\langle S_z \rangle = \frac{9}{2}\frac{G_A}{1+F/D} \left [ \frac{F}{D} - 
\frac{5}{9} \right ], \quad\quad (\mathrm{\chi QSM}).
\ee
In the limit $F/D=5/9$, which is the Skyrme model value, one has 
$\langle S_z \rangle = 0$, while in the NQM limit $G_A=\frac{5}{3}$, $F/D = 
\frac{2}{3}$ one has $\langle S_z \rangle = \frac{1}{2}$. For the measured values 
$F=0.46$, $D=0.80$, this model yields $\langle S_z \rangle =0.07$,
which is very close to the estimate in eq.(\ref{sz_lz_result}).

Information about $\langle S_z \rangle$ has also been derived from the
analysis of structure functions $g_1^{p,n}$ measured in polarized deep inelastic
scattering \cite{Adams,Abe}. The integrals of these structure functions can be
written as 
\be
\int\,g_1^{p,n}(x,Q^2)\,dx = \frac{C_1^{NS}(Q^2)}{12} \left [ \pm a^{(3)} + 
\frac{1}{3} a^{(8)} \right ] + \frac{C_1^{S}(Q^2)}{9} a^{(0)}(Q^2)
\ee
where $C_1^{NS}$ and $C_1^S$ are perturbatively calculable coefficients. 
The singlet axial coupling $a^{(0)}(Q^2)$ differs 
from $\Delta\Sigma=\Delta u +\Delta d+\Delta s$ as a consequence of the gluon
anomaly. In the Adler-Bardeen factorization scheme, $a^{(0)}(Q^2)$ is related to 
$\Delta\Sigma$ by
\be
a^{(0)}(Q^2) = \Delta\Sigma - n_f\frac{\alpha_s(Q^2)}{2\pi} \Delta G(Q^2)
\ee
where $\Delta G(Q^2)$ is the net polarization of gluons in a polarized nucleon.
A determination of $\Delta\Sigma$ from the measured quantity $a^{(0)}(Q^2)$
is only possible by invoking a model for the polarized gluon density, and 
fitting it 
to the observed $Q^2$-dependence of the structure functions. The result of one
such fit \cite{Altarelli} is 
\be
\langle S_z \rangle = \frac{1}{2}\Delta\Sigma = 0.22 \pm 0.045 \qquad
(\mathrm{polarized\,\,structure\,\,functions})
\ee
Other analyses (\cite{Abe},\cite{Adams},\cite{Reya}) obtain values of 
$\langle S_z \rangle$ between $0.1$ and $0.3$.
Within errors, the result for $\langle S_z \rangle$ obtained from high energy
experiments is compatible with the result (\ref{sz_lz_result}) 
obtained from a fit to the static properties.

It remains to be seen whether a specific test of rotational angular momentum
$\langle L_z \rangle$ and its radial distribution can be devised. We have
argued that the difference in shapes of the axial vector and isovector
magnetic form factors is a probe of orbital angular momentum. A precise
determination of $F_A(Q^2)$, which does not presume a dipole behaviour from
the outset, would be of great interest in this respect.



\newpage

\begin{center}
{\large Figure Captions}
\end{center}

\begin{itemize}
\item[Fig.1.] Flux tube connecting three constituent quarks, rotating
collectively around proton spin axis.
\item[Fig.2.] The fitted domain of $\langle S_z \rangle$ and $\langle L_z 
\rangle$ for the central value of $\mu_u$. The dotted lines represent the baryon
magnetic moments of table \ref{tab1}. 
\item[Fig.3.] Allowed domain of $\langle S_z \rangle$ and $\langle L_z \rangle$
for the full interval of $\mu_u = 2.16 \pm 0.08$.
\item[Fig.4.] The fitted domain of $f_{\mathrm{spin}}$ and $f_{\mathrm{orb}}$ 
for the central value of $\langle S_z \rangle$. The dotted lines represent 
the baryon magnetic moments of table \ref{tab1}. 
\item[Fig.5.] Predicted shape of axial vector form factor with
$f_{\mathrm{spin}} = 0.90$, $f_{\mathrm{orb}} = 0.10$, compared to a 
dipole with $M_A = 0.92$ GeV.
\end{itemize}

\vspace*{1cm}

\begin{center}
{\large Table Caption}
\end{center}

\begin{itemize}
\item[Table I.] Parametrization of magnetic moments in the rotating 
flux-tube model (model (A) of Ref. \cite{Casu}). The fits are based
on $\lambda = m_d/m_s = 0.6$, $\mu_u = -2\mu_d$, $\mu_s = 0.6\mu_d$
(for these values, the orbital contribution to the neutral baryons
$n$, $\Xi^0$ and $\Lambda^0$ vanishes).
\end{itemize}


\newpage

\begin{table}[b]
\begin{center}
\caption{}
\label{tab1}
\begin{tabular}{l}
$\mu(p) = \mu_u \Delta u + \mu_d \Delta d + \mu_s \Delta s  + \left [ 
\frac{2}{3}\mu_u + \frac{1}{3}\mu_d \right ] \langle L_z \rangle$ \\
$\mu(n) = \mu_u \Delta d + \mu_d \Delta u + \mu_s \Delta s  + \left [ 
\frac{1}{3}\mu_u + \frac{2}{3}\mu_d \right ] \langle L_z \rangle$ \\ 
$\mu(\Sigma^+) = \mu_u \Delta u + \mu_d \Delta s + \mu_s \Delta d  + \left [ 
\frac{2\lambda}{1+2\lambda}\mu_u + \frac{1}{1+2\lambda}\mu_s \right ] \langle 
L_z \rangle$ \\
$\mu(\Sigma^-) = \mu_u \Delta s + \mu_d \Delta u + \mu_s \Delta d  + \left [ 
\frac{2\lambda}{1+2\lambda}\mu_d + \frac{1}{1+2\lambda}\mu_s \right ] \langle 
L_z \rangle$ \\
$\mu(\Xi^-) = \mu_u \Delta s + \mu_d \Delta d + \mu_s \Delta u  + \left [ 
\frac{\lambda}{2+\lambda}\mu_d + \frac{2}{2+\lambda}\mu_s \right ] \langle 
L_z \rangle$  \\
$\mu(\Xi^0) = \mu_u \Delta d + \mu_d \Delta s + \mu_s \Delta u  + \left [ 
\frac{\lambda}{2+\lambda}\mu_u + \frac{2}{2+\lambda}\mu_s \right ] \langle 
L_z \rangle$  \\
$\mu(\Lambda^0) = \frac{1}{6} \left ( \Delta u + 4\Delta d + \Delta s \right ) 
\left ( \mu_u+\mu_d \right ) + \frac{1}{3} \left ( 2\Delta u - \Delta d + 
2\Delta s \right )\mu_s$ \\
$\,\,\,\,\,\,\,\,\,\,\,\,\,\,\,\,\,\,\,\,\,\,\,+ \left [ \frac{\lambda}
{1+2\lambda}\mu_u + 
\frac{\lambda}{1+2\lambda}\mu_d + \frac{1}{1+2\lambda}\mu_s \right ] \langle 
L_z \rangle$ 
\end{tabular}
\end{center}
\end{table}


\newpage

\newpage

\newpage    

\newpage

\newpage

\end{document}